\documentclass[aps, prx, notitlepage, citeautoscript, superscriptaddress, reprint]{revtex4-2}

\usepackage{xr}
\usepackage{amsmath}
\usepackage{amsfonts,amssymb,amsthm,amsxtra,physics,dsfont}
\usepackage[]{graphicx}
\usepackage{grffile}
\usepackage{mathrsfs}
\usepackage{framed}
\usepackage{bbm}
\usepackage{bm}
\usepackage{braket}
\usepackage{xcolor}
\usepackage{mathtools}
\usepackage[bookmarks=false,colorlinks=true,urlcolor=blue,citecolor=blue,linkcolor=blue]{hyperref}
\usepackage{algorithm}  
\usepackage{algpseudocode}
\usepackage{makecell}
\usepackage{lipsum}

\newcommand{\bth}{\boldsymbol{\theta}}

\newcommand{\av}[1]{\langle #1 \rangle}

\providecommand{\ket}[1]{\vert#1\rangle}
\providecommand{\bra}[1]{\langle#1\vert}

\newcommand{\be}{\begin{equation}}
\newcommand{\ee}{\end{equation}}
\newcommand{\bea}{\begin{eqnarray}}
\newcommand{\eea}{\end{eqnarray}}
\newcommand{\ba}{\begin{eqnarray*}}
\newcommand{\ea}{\end{eqnarray*}}

\definecolor{wrongultramarine}{rgb}{1,0.5,0}

\definecolor{forestgreen}{rgb}{0.13, 0.55, 0.13}

\renewcommand{\Re}{\text{Re}}
\renewcommand{\Im}{\text{Im}}

\newcommand{\bfvv}{{\boldsymbol{V}}}

\newcommand{\pder}[2]{\frac{\partial #1}{\partial #2}}
\newcommand{\round}[1]{\left( #1 \right)}
\newcommand{\curly}[1]{\left\{#1\right\}}
\renewcommand{\square}[1]{\left[ #1 \right]}
\newcommand{\nn}{\nonumber}
\DeclareMathOperator{\SD}{\mathcal{D}}

\def\bmx{\begin{pmatrix}}
\def\emx{\end{pmatrix}}

\begin{document}

\title{Sampling Noise and Optimized Measurement Distribution in Imaginary-Time Quantum Dynamics Simulations}

\author{Feng Zhang}
\email{fzhang@ameslab.gov}
\affiliation{Ames National Laboratory, Ames, Iowa 50011, USA}
\affiliation{Department of Physics and Astronomy, Iowa State University, Ames, Iowa 50011, USA}

\author{Niladri Gomes}
\affiliation{Ames National Laboratory, Ames, Iowa 50011, USA}

\author{Joshua Aftergood}
\affiliation{Ames National Laboratory, Ames, Iowa 50011, USA}
\affiliation{Department of Physics and Astronomy, Iowa State University, Ames, Iowa 50011, USA}

\author{Thomas Iadecola}
\affiliation{Ames National Laboratory, Ames, Iowa 50011, USA}
\affiliation{Department of Physics and Astronomy, Iowa State University, Ames, Iowa 50011, USA}
\affiliation{Department of Physics, Pennsylvania State University, University Park, PA 16802, USA}
\affiliation{Institute for Computational and Data Sciences, The Pennsylvania State University, University Park, Pennsylvania 16802, USA}
\affiliation{Materials Research Institute, The Pennsylvania State University, University Park, Pennsylvania 16802, USA}

\author{Yongxin Yao}
\email{ykent@iastate.edu}
\affiliation{Ames National Laboratory, Ames, Iowa 50011, USA}
\affiliation{Department of Physics and Astronomy, Iowa State University, Ames, Iowa 50011, USA}

\author{Peter P.~Orth}
\email{peter.orth@uni-saarland.de}
\affiliation{Ames National Laboratory, Ames, Iowa 50011, USA}
\affiliation{Department of Physics, Saarland University, 66123 Saarbr\"ucken, Germany}
\affiliation{Center for Quantum Technologies (QuTe), Saarland University, 66123 Saarbr\"ucken, Germany}

\begin{abstract}
Variational quantum dynamics simulations (VQDS) provide a promising route to simulate real- and imaginary-time quantum dynamics on noisy intermediate-scale quantum devices using fixed-depth circuits. However, their practical performance is strongly limited by sampling noise arising from a finite number of circuit measurements. In this work, we systematically investigate the impact of sampling noise on VQDS, with a focus on ground-state preparation in one-dimensional Ising spin models using imaginary time evolution. We compare different regularization strategies for stabilizing the equations of motion and show that Tikhonov regularization provides robust performance in noisy imaginary-time evolution. We then benchmark measurement-distribution strategies that allocate shots by minimizing a cost function that characterizes the error in solving the equation of motion. Using noisy circuit simulations, we demonstrate that such optimized shot allocation can significantly improve state fidelity and reduce the total measurement cost by more than a factor of two compared to uniform shot distributions. We observe that the best results are found if a sufficiently large number of measurements is guaranteed for all circuits, suggesting that a finite fraction of shots should be distributed evenly. Our results provide practical guidelines for implementing measurement-efficient variational quantum dynamics and ground-state preparation on near-term quantum hardware.
\end{abstract}

\date{\today}

\maketitle

\section{Introduction}
\label{sec:introduction}
Simulating the dynamics of quantum many-body systems is one of the most promising applications of quantum computing. While classical implementations of this problem generically require exponential resources to store and manipulate a volume-law entangled quantum state~\cite{sandvik_review,White92,Prosen2007}, there exist efficient implementations on a quantum computer~\cite{lloyd1996,childs2012hamiltonian,lowOptimalHamiltonianSimulation2017,lowHamiltonianSimulationQubitization2019,haahQuantumAlgorithmSimulating2021a}. For Hamiltonians containing only local interactions, the required resources for Hamiltonian simulation scale polynomially in the final simulation time $T$ and the number of qubits $N$~\cite{lowHamiltonianSimulationQubitization2019,childsFirstQuantumSimulation2018}. Methods that exhibit the best asymptotic scaling such as qubitization~\cite{lowHamiltonianSimulationQubitization2019} or quantum walk methods~\cite{berryHamiltonianSimulationNearly2015}, however, suffer from significant overhead costs in the required number of entangling gates and ancilla qubits.

In the current era of noisy intermediate-scale quantum (NISQ) computers~\cite{nisq}, the number of qubits is still limited, and the maximal circuit depth is bounded by a finite device coherence time. This severely limits the direct application of Trotter product formulas and other even more resource-intensive algorithms. Trotter dynamics simulations can thus currently only reach final simulation times $T$ that are of order one to ten in units of the natural energy scale of the Hamiltonian~\cite{Martinez_2016, Lamm_2018, Trotter_dynamics_Knolle,freyRealizationDiscreteTime2022a,kim2023scalable,Chen2022,kimEvidenceUtilityQuantum2023, chen2023problemts, chertkovRobustnessNearthermalDynamics2026}. This has motivated research on variational quantum dynamics simulations with fixed-depth parameterized quantum circuits, applicable along both the real-time~\cite{Li_Benjamin-PRX-2017,theory_vqs,AVQDS,lin2021real,barrattParallelQuantumSimulation2021,barison2021efficient,benedetti2020hardware,mansurogluClassicalVariationalOptimization2021,berthusenQuantumDynamicsSimulations2022}, and the imaginary-time axis~\cite{VQITE, theory_vqs}, the latter alternatively referred to as variational quantum imaginary-time evolution (VQITE). One example is VQDS based on McLachlan's variational principle~\cite{Li_Benjamin-PRX-2017,theory_vqs}, where one propagates the parameters $\boldsymbol{\theta}$ of a variational wavefunction $\ket{\Psi(\boldsymbol{\theta}(t))}$ 
in real (or imaginary) time according to \emph{classical} equations of motion $M \dot{\boldsymbol{\theta}} = V$. The coefficients $M_{\mu\nu}$ and $V_\mu$ are calculated on a quantum computer. Recent work has aimed to reduce the measurement overhead of the quantum geometric tensor (QGT) $M$, which exhibits unfavorable quadratic scaling with respect to the number of variational parameters~\cite{gacon2023ieee, Gacon2024VariationalQTE, Anuar2024OperatorPV}. For example, operator-projected VQITE relaxes the requirement that the full variational state follows the imaginary-time evolution path, instead enforcing that a selected set of operator expectation values satisfy the projected imaginary-time equations. This approach circumvents explicit QGT estimation and correspondingly reduces the measurement cost scaling with the number of variational parameters~\cite{Anuar2024OperatorPV}.

VQDS with an adaptive ansatz~\cite{AVQDS} (AVQDS) was shown to exhibit a favorable linear scaling of the number of parameters with early simulation times $N_\theta \propto T$ when describing generic nonequilibrium real-time dynamics of nonintegrable spin models such as the mixed-field Ising model (MFIM). The number of parameters $N_{\theta}$ was found to saturate at long times to a value $N^{\text{sat}}_{\theta}$. Numerical benchmarks for nonintegrable models indicate that $N^{\mathrm{sat}}_{\theta}$ grows exponentially with system size $N$. In contrast, for the integrable transverse-field Ising model (TFIM), a polynomial scaling $N_\theta \propto N^{2}$ is observed, consistent with fast-forwarding algorithms~\cite{atiaFastforwardingHamiltoniansExponentially2017,cirstoiu2020variational,gibbsLongtimeSimulationsHigh2021,kokcuAlgebraicCompressionQuantum2022,bassmanConstantdepthCircuitsDynamic2022,campsAlgebraicQuantumCircuit2021,gulaniaQuantumTimeDynamics2021}. This makes VQDS most attractive for simulating early-time dynamics of general models and long-time dynamics for integrable systems, where one expects a polynomial scaling of $N_{\theta}$ with $T$ and $N$.The imaginary-time version of AVQDS, termed Adaptive VQITE (AVQITE), was shown to produce compact ground-state ans\"atze for molecules and spin models~\cite{AVQITE}, with performance comparable to other adaptive ansatz construction methods~\cite{grimsleyAdaptiveVariationalAlgorithm2019, QubitAdaptVQE,Yordanov2021Qubitea}. Additional applications include finite-temperature simulations~\cite{getelina2023adaptive, chen2025minimally}, calculations of Green’s functions and nonlinear response functions~\cite{mootz2023twodimensional, Gomes2023, mootz2024adaptive}, evaluations of topological properties~\cite{Mootz2026EfficientBP}, and simulations of systems governed by non-Hermitian Hamiltonians, including open quantum systems~\cite{Chen2024_avqite_open} and certain formulations of partial differential equations~\cite{alipanah2025quantum}.

Previous studies of VQDS have mostly relied on statevector quantum simulations, which correspond to a perfect quantum computer using an infinite number of measurements to obtain its results. Here, we explore in detail the impact of shot noise arising from a finite number $\mathcal{M}$ of measurements in VQDS. We consider state propagation of one-dimensional quantum spin models, focusing on imaginary time, with a particular emphasis on ground-state preparation. This is a well-suited task for VQDS due to the expectation that ground states with area-law entanglement entropy in one dimension can be prepared by circuits of modest depth~\cite{Hastings07,Eisert10,Schollwock11}. It was previously found that sampling noise due to finite $\mathcal{M}$ can have a substantial detrimental effect on the performance of VQDS, since the required inversion of the often ill-conditioned matrix $M$ leads to an unfavorable scaling of $\mathcal{M}$ with the condition number~\cite{theory_vqs,VQITE,stokes2020quantum,benedetti2020hardware,gacon2023ieee}. We show that one can alleviate this issue by properly regularizing the matrix.

We also address the important question of how to best distribute the number of measurements over the various quantum circuits that need to be evaluated to obtain the elements $M_{\mu\nu}$ and $V_\mu$. Within the context of the variational quantum eigensolver (VQE), several schemes to optimize the distribution of measurements have been discussed recently~\cite{Sweke2020stochasticgradient,Kubler2020adaptiveoptimizer, Crawford2021efficientquantum, arrasmith2020operator, greschGuaranteedEfficientEnergy2025, Eckstein-PRX_Q-2026}. More specific to VQDS, it was recently proposed in Ref.~\cite{PRXQuantum.2.030324} to distribute measurements among the elements in a way that minimizes the variance in the solution to the equations of motion. Here, we systematically benchmark this strategy on noisy quantum simulators. We demonstrate that implementing a minimal number of shots is necessary for this strategy to be advantageous against uniform shot distribution. Moreover, we also compare the performance of several alternative definitions of the cost function.

The remainder of the article is organized as follows. In Sec.~\ref {sec:algo}, we introduce the VQDS algorithms, the variational ansatz we use and the spin model we study. The effects of sampling noise will be discussed in Sec.~\ref{sec:effects_sampling_noise}, where we systematically benchmark strategies for regularizing the quantum geometric tensor, as well as for allocating shots onto the different quantities that define the equation of motion of the variational parameters, in order to alleviate the shot noise. Finally, we summarize our work in Sec.~\ref{sec:conclusions}.  

\section{Algorithms, ansatz and model}
\label{sec:algo}
\subsection{VQITE algorithm}
\label{subsec:VQDS_VQITE_algo}
Imaginary time evolution is described by the imaginary-time von Neumann equation~\cite{VQITE,beach2019making,qite_chan20,gomesAdaptiveVariationalQuantum2021}
\begin{equation}
\frac{d \rho}{d\tau} = \mathcal{L}(\rho) = - \bigl\{ H, \rho \bigr\} + 2 \av{H} \rho \,.
\end{equation}
Here, $\rho(\tau)$ denotes the density matrix and $\tau$ is the imaginary-time parameter. Furthermore, $\{H,\rho\} = H\rho + \rho H$ denotes the anticommutator, and $\av{H}(\tau) = \mathrm{Tr}[H\rho(\tau)]$ is the energy expectation value.
 As long as the initial state $\rho(0)$ has finite overlap with the ground-state $\rho_{\rm GS}$, $\rho(\tau)$ is guaranteed to converge to the ground state in the large-$\tau$ limit:
\begin{equation}
\rho_{\text{GS}} = \lim_{\tau \rightarrow \infty} \frac{e^{- H \tau} \rho(0) e^{-H \tau}}{\text{Tr}(e^{-2 H \tau} \rho(0))} \,.
\end{equation}

For completeness, we now briefly describe the main steps required to implement the imaginary-time (VQITE) variational dynamics based on McLachlan's principle~\cite{mclachlan64variational}. For further details, we refer the reader to Ref.~\onlinecite{theory_vqs} (see also Refs.~\cite{AVQDS,gomesAdaptiveVariationalQuantum2021}). We note that variational real-time dynamics can be simulated in a similar fashion. The quantum circuits that need to be measured have the same structure for real- and imaginary time VQDS such that our regularization and shot distribution methods described in Sec.~\ref{sec:effects_sampling_noise} can equally be applied to real-time VQDS. 

First, we choose a variational parameterization of the time-evolving state, $\ket{\Psi(\boldsymbol{\theta})}$, where $\boldsymbol{\theta}$ is a vector of $N_{\bth}$ real parameters. The variational ansatz needs to be sufficiently expressive to represent the exact time-evolved state with high fidelity. Under this assumption, and assuming a sufficiently small propagation time step, it is guaranteed by the variational principle that $\ket{\Psi[\boldsymbol{\theta}(t)]}$ follows the exact time evolution with high fidelity. The time evolution of the variational parameters is governed by a classical equation of motion (EOM)
\be
\sum_\nu M_{\mu \nu}\dot{\theta}_\nu = V_\mu \,.
\label{eq:eom}
\ee
This equation is derived by minimizing the McLachlan distance $L^2$ between the exact and the variational time evolution over a single time step:
\bea
L^2 &\equiv&\norm{\sum_\mu \frac{\partial \rho[\bth]}{\partial \theta_\mu} \dot{\theta}_\mu - \mathcal{L}[\rho]}^2 \notag \\
&=&\sum_{\mu, \nu} M_{\mu \nu} \dot{\theta}_\mu \dot{\theta}_\nu -2 \sum_\mu V_\mu \dot{\theta}_\mu + 2 \text{var}_{\bth}[H] \,.
\label{L2}
\eea
Here, $\norm{A} = \sqrt{\text{Tr}(A^\dag A)}$ is the Frobenius matrix norm and $\text{var}_{\bth}[H] = \av{H^2}_{\bth} - \av{H}_{\bth}^2 $ is the variance of the Hamiltonian in the variational state. The explicit form of the matrix $M_{\mu \nu}$ and the vector $V_\mu$ for imaginary time evolution is given by 
\begin{align}
    M_{\mu \nu} &= \Re\biggl[\frac{\partial \bra{\Psi[\bth]}}{\partial \theta_\mu} \frac{\partial \ket{\Psi[\bth]}}{\partial \theta_\nu} \nonumber \\ & + \bra{\Psi[\bth]} \frac{\partial \ket{\Psi[\bth]}}{\partial \theta_\mu} \bra{\Psi[\bth]} \frac{\partial \ket{\Psi[\bth]}}{\partial \theta_\nu}\biggr] \\
    V_\mu &= - \Re\biggl[\frac{\partial\bra{\Psi}}{\partial \theta_\mu}H\ket{\Psi}\biggr] \,.
\end{align}
We note that in order to simulate real-time VQDS one only needs to change the form of the gradient to $V_\mu^{\mathrm{(real-time)}} = -\Im\Bigl[\frac{\partial\bra{\Psi}}{\partial \theta_\mu}\ket{\Psi}\ev{H}{\Psi}
 -\frac{\partial\bra{\Psi}}{\partial \theta_\mu}H\ket{\Psi} \Bigr]$. The numerical values of $M_{\mu \nu}$ and $V_\mu$ can be obtained by executing quantum circuits; see e.g.~Ref.~\cite{AVQDS} for an explicit form of the circuits. Given a finite number of circuit evaluations, the elements of $M$ and $V$ exhibit uncertainty due to sampling noise that decreases as $1/\sqrt{\mathcal{M}_{\alpha}}$, where $\mathcal{M}_\alpha$ denotes the number of measurements of a particular element of $M$ or $V$. In Sec.~\ref{sec:effects_sampling_noise}, we discuss an optimized way of distributing the total number of measurements over each element (compared to a uniform distribution) assuming a fixed total measurement budget.

\subsection{Models and variational ansatz}
\label{subsec:models_ansatz}
We apply the VQITE algorithms to the $N$-site one-dimensional (1D) transverse-field Ising model (TFIM) with open boundary conditions:
\begin{equation}
H_{\text{TFIM}} = -J \sum_{j=1}^{N-1} Z_{j}Z_{j+1} - \Delta \sum_{j=1}^N X_{j} \,.
\label{eq:tfim}
\end{equation}
We focus on TFIM with parameters $\Delta = J = 1$, which corresponds to the critical point separating PM and FM phases in the thermodynamic limit.

\begin{figure}[tb]
    \centering
    \includegraphics[width=\linewidth]{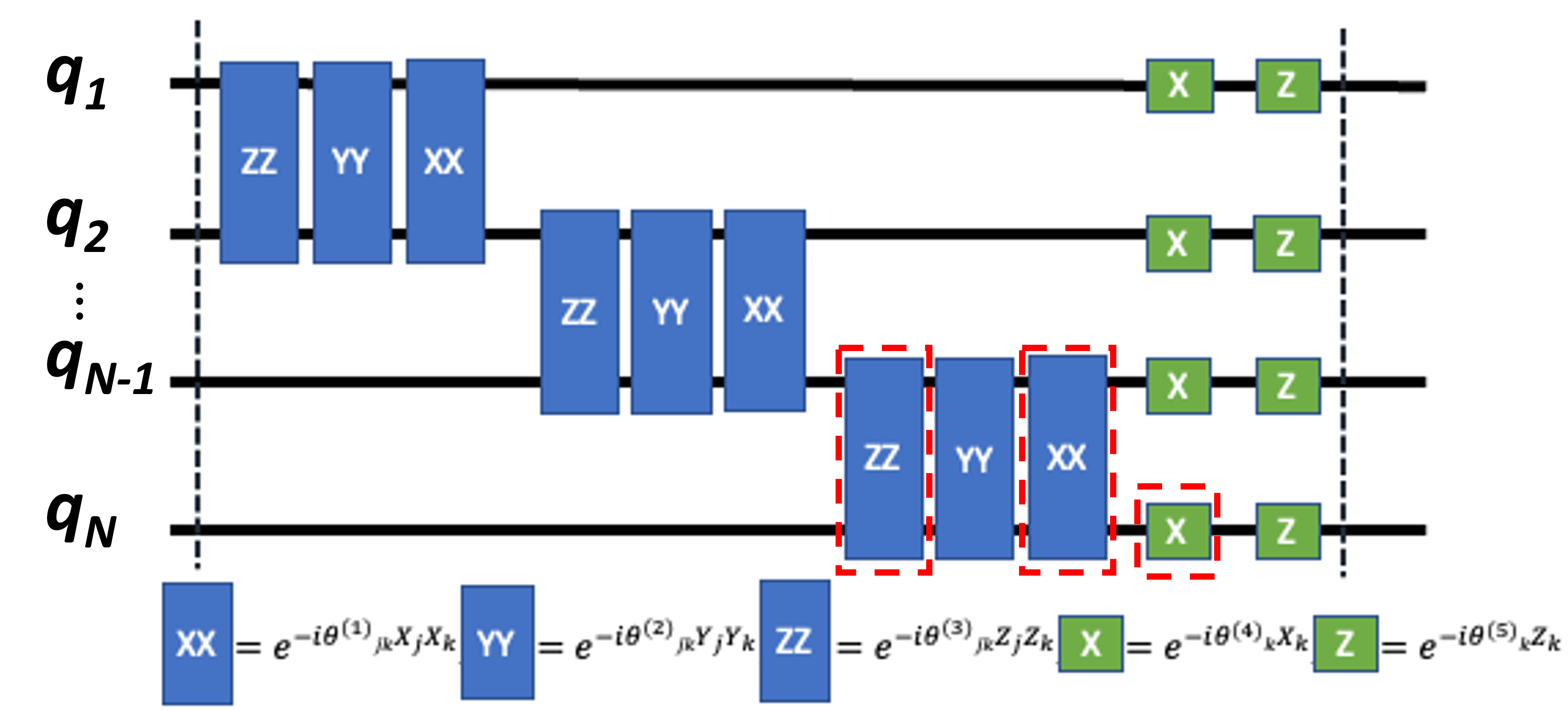}
    \caption{Schematic illustration of one layer of parametrized unitaries in the variational ansatz~\eqref{eq:ansatz}. The blue boxes represent the two-qubit operators that are ordered sequentially to allow for unrestricted qubit correlations across a single layer. The single-qubit rotation gates are shown in green boxes. Each rotation gate has its own variational parameter. The dashed red boxes indicate the three operators that are not included in the final ansatz to remove an apparent singularity in the QGT $M_{\mu \nu}$.}
    \label{fig:qcircuit}
\end{figure}

For the variational wavefunction, we use a layered ansatz of the form shown in Fig.~\ref{fig:qcircuit}:
\begin{equation}
\ket{\Psi(\bth)} = \prod_{\ell = 1}^p \prod_{j = 1}^{N_\theta/p} e^{-i \theta_{\ell, j} P_j}\ket{\psi_0}\,.
\label{eq:ansatz}
\end{equation}
Here, each $P_j$ is a single Pauli string that we choose to act nontrivially only on a single qubit or on a nearest-neighbor pair of qubits. This ensures that this ansatz can be efficiently implemented on quantum devices with nearest-neighbor connectivity. We arrange the gates in sequential form as shown in Fig.~\ref{fig:qcircuit}, which depicts a single layer of the circuit. The sequential ordering of the gates has the advantage that correlations can spread across all qubits even for a single layer, and the correlation length of the variational states is not restricted~\cite{lin2021real,barrattParallelQuantumSimulation2021,smithCrossingTopologicalPhase2022,dborinSimulatingGroundstateDynamical2022}.

In practice, we set the reference state $\ket{\psi_0}$ in Eq.~\eqref{eq:ansatz} as a product state $\bigotimes_{i=1}^{N}\ket{+}_i$,  in which $\ket{+}_i$ is the eigenstate of  $X_i$ with the eigenvalue of 1. As a result of this choice of the reference state and the particular ordering of the operators as shown in Fig.~\ref{fig:qcircuit}, the matrix $M$ is always singular. In particular, the following features of $M$ always hold regardless of the rotation angles on the operators: (1) the two columns corresponding to Pauli strings $Y_{N}Y_{N-1}$ and $Z_{N}Z_{N-1}$ are anti-parallel; (2) the two columns corresponding to $X_{N-1}Y_{N-2}$ and $X_{N}$ are parallel; and (3) the column corresponding to $X_{N}X_{N-1}$ is zero. For this reason, we remove three operators, namely, $Z_NZ_{N-1}$, $X_N$, and $X_NX_{N-1}$, in the following calculations.

\section{Effects of sampling noise}
\label{sec:effects_sampling_noise}
\subsection{Inversion of noisy matrix M}
\label{subsec:inverting_M}
The time evolution of the variational parameters $\bth(t)$ is obtained by integrating the equation of motion~\eqref{eq:eom}. Choosing the Euler method, one finds that the change of $\bth(t)$ during a time step $\delta t$ reads
\be
\delta \bth = M^{-1} \bfvv \delta t \,.
\label{eq:theta}
\ee
Note that $M_{\mu\nu}$ and $V_\mu$ depend on the variational parameters $\bth(t)$. The global truncation error over the total simulation period scales linearly with the step size $\delta t$, and we set a maximal step size $\delta t_{\text{max}}$ when executing the algorithm. The change of the wavefunction $\ket{\Psi(\bth)}$ in one step, however, is determined by $\delta \bth$ and thus depends on both $\delta t$ and $\dot{\bth} \equiv M^{-1}V$. Rather than working with a fixed step size $\delta t$, we therefore impose a maximally allowed parameter change $\delta \theta_{\text{max}}$. We then vary $\delta t$ in each time step accordingly such that $|\delta \theta_\mu| \leq \delta \theta_{\text{max}} \forall \mu$ and $\delta t \leq \delta t_{\text{max}}$.

Even after the removal of certain operators in the ansatz that cause singularities in the QGT $M$ as described above, the condition number of $M$ can still become large, defined as $\kappa = \text{max}_i(\abs{\lambda_i})/\text{min}_i(\abs{\lambda_i})$, where $\lambda_i$ are the eigenvalues of $M$. This makes the direct matrix inversion in Eq.~\eqref{eq:theta} particularly challenging in the presence of shot noise. It was shown previously that a large condition number can increase the required number of measurements by a factor of $\kappa$ in order to keep the uncertainty of $\delta \bth$ below a fixed threshold~\cite{benedetti2020hardware}.

\begin{figure}[t]
    \centering
    \includegraphics[width=\linewidth]{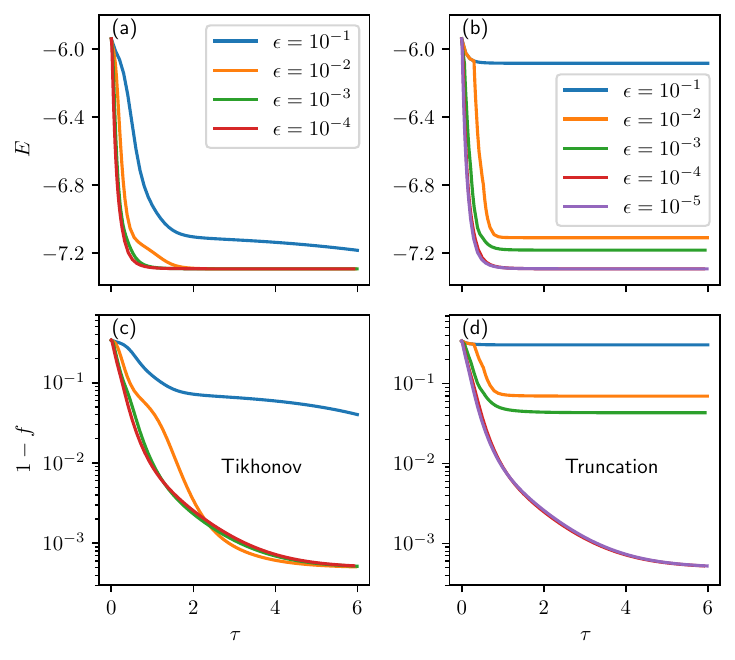}
    \caption{Energy $E$ and infidelity $1 - f$ of noiseless VQITE for the critical TFIM with $N=6$ spins. Tikhonov regularization [(a) and (c)] and eigenvalue truncation [(b) and (d)] are used, respectively, to solve the equations of motion. 
 }
    \label{fig:regularize}
\end{figure}

Different methods have been proposed to regularize $M$. Here, we focus on two of them: Tikhonov regularization~\cite{Tikhonov43} and eigenvalue truncation~\cite{hansen1990trucated,gacon2023ieee}. These two methods have previously been investigated for simulating the real-time variational quantum dynamics~\cite{Zhang2025TETRIS}; here, we compare their performance on the imaginary-time dynamics. Tikhonov regularization involves adding a diagonal matrix, $M \rightarrow \tilde{M} = M + \epsilon\openone$, before inversion. In the eigenvalue truncation method, $M^{-1}$ is approximated by $\sum_{i, \lambda_i>\epsilon}\lambda_i^{-1}u_iu_i^\dagger$, where $u_i$ are the eigenvectors of $M$ with the eigenvalue $\lambda_i$ and the summation only includes the eigenvalues that are greater than a cut-off threshold $\epsilon$. To determine the acceptable values for the parameter $\epsilon$, we first run noiseless statevector simulations for the critical TFIM with $N=6$ using both methods. Figure~\ref{fig:regularize} shows the evolution of the energy and infidelity with various values of $\epsilon$  for either method.  The infidelity is calculated based on the overlap with the ground state, which is obtained using exact diagonalization.
\begin{equation}
1-f = 1-|\ip{\Psi[\bth(\tau)]}{\Psi_{\text{GS}}}|^{2}\,.
\end{equation} 
For Tikhonov regularization, the variational state converges to the ground state for $\epsilon \leq 10^{-2}$ [Fig.~\ref{fig:regularize} (a) and (c)], while $\epsilon \leq 10^{-4}$ needs to be used for the eigenvalue truncation method [Fig.~\ref{fig:regularize} (b) and (d)]. 
Since too small values of $\epsilon$ can easily be overwhelmed by the noise, the eigenvalue truncation method is expected to have less tunability than the Tikhonov regularization method. 

\begin{figure}[t]
    \centering
    \includegraphics[width=\linewidth]{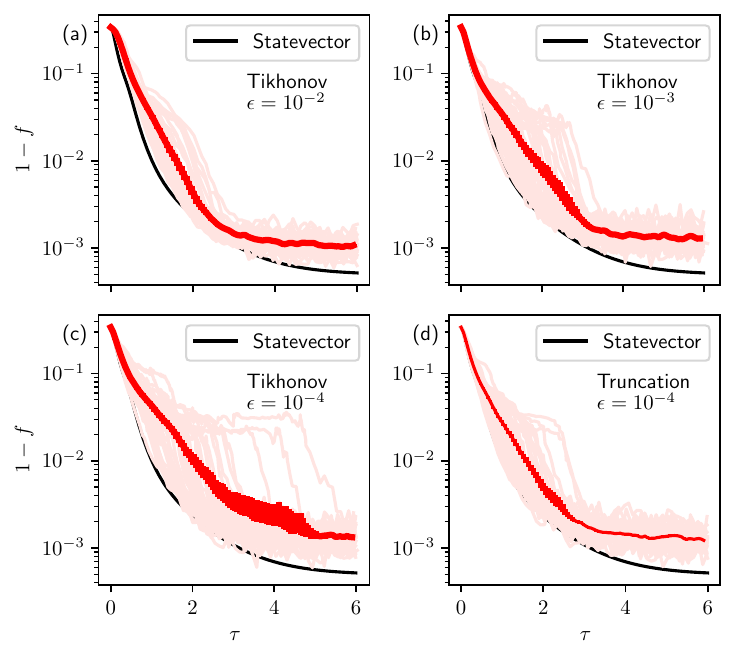}
    \caption{Wavefunction infidelity $1-f$ as a function of imaginary time $\tau$ in circuit simulations considering shot noise using Tikhonov regularization with $\epsilon=10^{-4}$ (a), $10^{-3}$ (b) and $10^{-2}$ (c), and using the eigenvalue truncation method with $\epsilon=10^{-4}$ (d). We use $\mathcal{M} = 10^4$ shots for each circuit measuring $D_{\mu \nu}, \mathcal{O}_\mu$ and $E_{\mu}^{\pm}$ for a total shot budget of about $4\times 10^6$ in each time step. 
    Each figure shows the overlay of $50$ independent runs in light red. The solid red line gives the average over the $50$ runs.
 }
    \label{fig:regularize_noise}
\end{figure}

To verify this expectation, we consider the effect of shot noise in simulations, where quantum circuits are constructed to measure $M$ and $V$: 
$M_{\mu\nu}=\Re(\mathcal{D}_{\mu\nu}+\mathcal{O}_\mu\mathcal{O}_\nu)$ and $V_\mu=-\Re(\mathcal{V}_\mu)$, where $\mathcal{D}_{\mu \nu}=\pdv{\bra{\Psi}}{\theta_{\mu}}\pdv{\ket{\Psi}}{\theta_{\nu}}$, $\mathcal{V}_\mu=\frac{\partial\bra{\Psi}}{\partial \theta_\mu}H\ket{\Psi}$, $\mathcal{O}_{\mu}=\pdv{\bra{\Psi}}{\theta_{\mu}}\ket{\Psi}$. Each element of $\mathcal{D}_{\mu\nu}$, and $\mathcal{O}_\mu$ is measured using a unique quantum circuit that exploits the Hadamard test method~\cite{AVQDS}, while the real part of $\mathcal{V}_\mu$ can be measured using the parameter shift rule: $\Re(\mathcal{V}_\mu)=(E_\mu^+-E_\mu^-)/4$, where $E_\mu^\pm=\bra{\Psi(\theta_\mu\pm\frac{\pi}{2})}H\ket{\Psi(\theta_\mu\pm\frac{\pi}{2})}$. For the TFIM Hamiltonian, there are two terms in the total energy $E$, $E_Z=\bra{\Psi}-J\sum_{j=1}^{N-1} Z_{j}Z_{j+1}\ket{\Psi}$ and $E_X=\bra{\Psi}-\Delta\sum_{j=1}^N X_{j}\ket{\Psi}$. These can be measured directly in two distinct circuits, one in the $Z$-basis and the other in the $X$-basis, respectively. We consider shot noise arising from a finite number of shots for each raw measurement. Figure~\ref{fig:regularize_noise} shows the evolution of the infidelity with imaginary time during these simulations, using both the Tikhonov regularization method with $\epsilon=10^{-2}$, $10^{-3}$, and $10^{-4}$, and the eigenvalue truncation method with $\epsilon=10^{-4}$. A total number of $10^4$ shots were used for each measurement. 50 independent simulations were performed for each parameter set, and the trajectory of each run is shown individually in light red, with the solid dark red line indicating the average over the 50 runs. As shown in Fig.~\ref{fig:regularize_noise} (a)-(c), with the value of $\epsilon$ used in Tikhonov regularization decreasing, one can see a wider spread of the 50 independent trajectories, whose average also deviates away from the statevector results, indicating a worse performance of regulating the $M$ matrix.   For the eigenvalue truncation method [Fig.~\ref{fig:regularize_noise} (d)], with $\epsilon=10^{-4}$, while it appears to perform better than Tikhonov regularization with the same $\epsilon$ [Fig.~\ref{fig:regularize_noise} (c)],  its overall performance is still worse than Tikhonov regularization with $\epsilon=10^{-2}$ given in Fig.~\ref{fig:regularize_noise} (a). Based on these results, we will use the Tikhonov regularization method with $\epsilon=10^{-2}$ in the following calculations.

\subsection{Optimized measurement distribution}
\label{subsec:optimized_meas_dist}
Now we address the important question of how to distribute a fixed total measurement budget $\mathcal{M}_{\text{tot}} = \sum_{\kappa} \mathcal{M}_\kappa$ among the different elements in order to maximize the state evolution fidelity. In Ref.~\cite{PRXQuantum.2.030324}, the authors established a strategy of distributing the shots by minimizing a cost function defined as the total variance of $\dot{\bth}$:
\be
\sigma_{\dot{\bth}}^2 =  \sum_\mu\sigma_{\dot{\theta_\mu}}^2 
=  \sum_\mu\sum_\kappa\left|\frac{\partial\dot{\theta}_\mu}{\partial m_\kappa}\right|^2\frac{\sigma_{m_\kappa}^2}{\mathcal{M}_\kappa},
\label{eq:variance}
\ee
where $\{m_\kappa\}$ is the collection of all the quantities that need to be measured, including $\mathcal{D}_{\mu \nu}$, $\mathcal{O}_\mu$, $E_Z$ and $E_X$. Also, $\sigma_{m_\kappa}^2$ denotes the intrinsic variance of $m_\kappa$, which can be calculated as $\sigma_{m_\kappa}^2=\langle m_\kappa^2\rangle-\langle m_\kappa\rangle^2$, and $\mathcal{M}_\kappa$ is the number of measurements used to measure $m_\kappa$. The variance of the measured expectation value of $m_\kappa$ scales as $1/\mathcal{M}_\kappa$. In Eq.~\eqref{eq:variance}, we use the variance propagation rule to calculate $\sigma_{\dot{\bth}}^2$ based on those of the raw measurements, assuming all the raw measurements are independent and the individual variances are sufficiently small. In Appendix~\ref{appendix:a}, we provide the derivation of the analytical expressions for $\frac{\partial\dot{\theta}_\mu}{\partial m_\kappa}$.

Given a total number of shots (also referred to as total measurement budget), the goal is to determine a strategy that allocates the shots on each measurement such that $\sigma_{\dot{\bth}}^2$ is minimized. The total measurement budget is denoted as $\mathcal{M}_{\text{tot}}=\sum_\kappa\mathcal{M}_\kappa$. This optimization problem with constraints is easily solvable, leading to a solution in which each 
\begin{equation}
    \mathcal{M}_\kappa\propto p_\kappa\equiv\sqrt{\sum_{\mu}\left|\frac{\partial\dot{\theta}_\mu}{\partial m_\kappa}\right|^2}\sigma_{m_\kappa}\,.
\end{equation}  
However, we find that this results in  $\mathcal{M}_\kappa$ being too small in practice for some values of $\kappa$ if the corresponding $\frac{\partial\dot{\theta}_\mu}{\partial m_\kappa}$ is very small. Going beyond Ref.~\cite{PRXQuantum.2.030324}, we therefore introduce an additional constraint that $\mathcal{M}_\kappa\geq\mathcal{M}_\mathbf{min}$. This strategy is implemented in the recursive Algorithm~\ref{alg:shotAlloc}.

\begin{algorithm}[H]
    \caption{Algorithm for shot allocation.}\label{alg:shotAlloc}
    \begin{algorithmic}[1]
        \Function{allocateShots}{$p[1:n],\mathcal{M}_\mathbf{min},\mathcal{M}_\mathbf{tot}$}
        \State /* $p[1:n]$: list of $p_\kappa$ in ascending order. */
        \State /* $\mathcal{M}_{\mathbf{min}}$: minimal number of shots. */
        \State /* $\mathcal{M}_{\mathbf{tot}}$: total number of shots. */
        \State $\mathcal{M}[1:n]\gets 0$
        \State $\mathbf{count}\gets 0$
        \For{$\kappa=1$ $\textbf{to}$ $n$}
            \State $\mathcal{M}_\kappa\gets p_\kappa*\mathcal{M}_\mathbf{tot}/\textbf{sum}(p[1:n])$
            \If{$\mathcal{M}_\kappa<\mathcal{M}_\textbf{min}$}
                \State $\mathbf{count}\gets\mathbf{count}+1$
            \EndIf
        \EndFor
        \If{$\mathbf{count}=0$}
            \State\Return{$\mathcal{M}[1:n]$}
        \Else 
            \State $\mathcal{M}[1:\mathbf{count}]\gets\mathcal{M}_\mathbf{min}$
            \State \parbox[t]{200pt}{$\mathcal{M}[\mathbf{count}+1:n]\gets \Call{allocateShots}{p[\mathbf{count}+1:n], \mathcal{M}_\mathbf{min},\mathcal{M}_\mathbf{tot}-\mathbf{count}*\mathcal{M}_\mathbf{min}}$\strut}
            \State\Return{$\mathcal{M}[1:n]$}
        \EndIf
        \EndFunction
    \end{algorithmic}
\end{algorithm}

\begin{figure}[t]
    \centering
    \includegraphics[width=\linewidth]{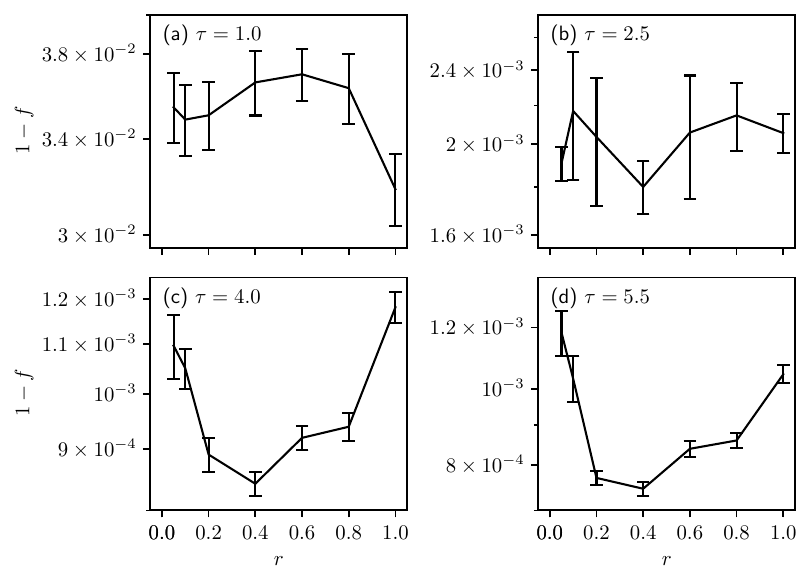}
    \caption{The infidelity $1-f$ as a function of $r$ for various times $\tau$ of noisy VQITE simulations of the critical TFIM with $N=6$. We set $\mathcal{M}_{\text{ave}} = 10^4$. For each $r$, which controls the number of minimal shots via $\mathcal{M}_\textbf{min}=r\mathcal{M}_\textbf{ave}$, the infidelity $1-f$ is averaged over $50$ independent runs with the error bar showing the error of the mean. We employ Tikhonov regularization with $\epsilon=10^{-2}$ to solve the equations of motion.
 }
    \label{fig:N6-f-r}
\end{figure}

In practice, we implement the constraints by setting an average number of shots for each measurement $\mathcal{M}_\text{ave}$, and letting $\mathcal{M}_\text{min}=r\mathcal{M}_\text{ave}$, where $0<r\leq 1$. If $r=1$, the shots are uniformly distributed among the different quantities $m_\kappa$. Let us now analyze the shot distribution strategy using simulations of the critical TFIM with system size $N=6$. 

In Fig.~\ref{fig:N6-f-r}, we plot the average $1-f$ as a function $r$ for a few time instants. $\mathcal{M}_\text{ave}$ is set to $10^4$. The infidelity is averaged over 50 independent runs for each $r$, with the error bar showing the error in estimating the average. The effectiveness of the shot allocation algorithm is not evident in the early stage at $\tau=1.0$, when the uniform distribution ($r=1$) has the lowest infidelity. On the other hand, an intermediate value, $r=0.4$, emerges as optimal in all the subsequent snapshots shown in Fig.~\ref{fig:N6-f-r}. At later stages with $\tau=4.0$ and $5.5$, where the state approaches the ground state, one can observe a noticeable improvement compared to the uniform distribution when choosing $r < 1$, except at $r=0.05$, which has the worst performance at the end of the simulation. This indicates that it is essential to maintain a sufficiently large number of minimal shots for each quantity $m_\kappa$.

\begin{figure}[t]
    \centering
    \includegraphics[width=\linewidth]{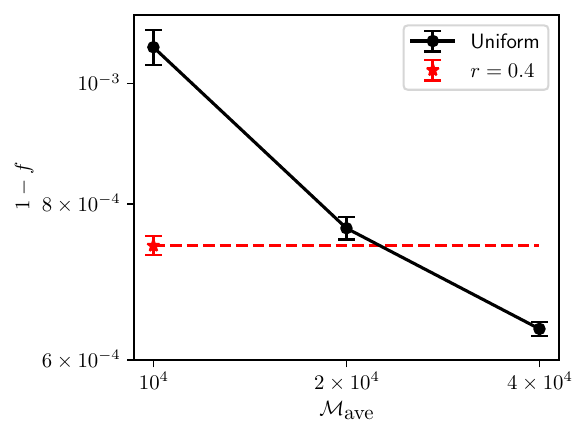}
    \caption{Infidelity $1-f$ as a function of $\mathcal{M}_\text{ave}$ for uniform shot distribution for critical TFIM with $N=6$ at $\tau=5.5$ (black points). The implementation of Algorithm~\ref{alg:shotAlloc} with $r=0.4$ and $\mathcal{M}_\text{ave}=10^4$ at the same $\tau$ is shown as the red star. The horizontal red dashed line indicates how many average measurements $\mathcal{M}_\text{ave}$ are needed in the uniform case to achieve the same infidelity, signaling a cost reduction by about 50\%. All data points are averaged over 50 independent runs.
 }
    \label{fig:uniform-cmp}
\end{figure}

To quantitatively show how many shots can be saved by implementing the shot distribution Algorithm~\ref{alg:shotAlloc}, we performed noisy simulations with uniform shot distribution for larger $\mathcal{M}_\textbf{ave}$. Again, we consider the critical TFIM and perform $50$ independent runs. Figure~\ref{fig:uniform-cmp} shows the averaged infidelity $1-f$ as a function of $\mathcal{M}_\text{ave}$. As expected, more accurate results are obtained when the total number of shots increases. Larger $\mathcal{M}_\text{ave}$ also helps with the convergence, as evidenced by the shrinking error-bar. On the other hand, when Algorithm~\ref{alg:shotAlloc} is implemented with $\mathcal{M}_\text{ave}=10^4$ and $r=0.4$ (red star), we find an infidelity that could only be achieved with $\mathcal{M}_\text{ave}>2\times10^4$ when using a uniform shot distribution. This corresponds to saving about 50\% of the total number of shots when using the optimized measurement distribution compared to the uniform case.  

\begin{figure}[t]
    \centering
    \includegraphics[width=\linewidth]{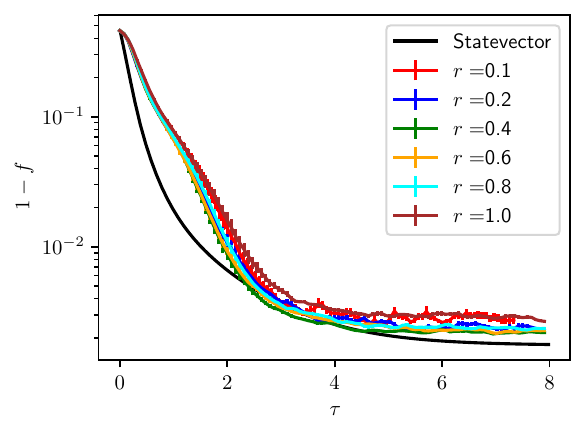}
    \caption{Infidelity $1-f$ as a function of the imaginary time $\tau$ for various values of $r$ in noisy VQITE simulations of the critical TFIM with system size $N=8$. We set $\mathcal{M}_\text{ave} = 10^4$. For each $r$, which controls the minimal number of shots as $\mathcal{M}_\text{min}=r\mathcal{M}_\text{ave}$, the infidelity $1-f$ is averaged over 20 independent runs. The error bar indicates the error of the mean. 
    We use Tikhonov regularization with $\epsilon=10^{-2}$ to solve the equations of motion. Noiseless statevector results are shown for comparison. 
 }
    \label{fig:N8-f-t}
\end{figure}

We also performed simulations on a larger system with $N=8$. Fig.~\ref{fig:N8-f-t} plots $1-f$, averaged over 20 runs for each $r$, as a function of the imaginary time $\tau$. Appreciable improvement in the wavefunction fidelity compared with the uniform distribution  ($r=1$) can be observed for all values of $r<1$ except for $r=0.1$, again, demonstrating the effectiveness of the shot allocation strategy and the necessity of implementing a minimal number of shots. Also, similar to the case of $N=6$, $r=0.4$ appears to be an optimal value for achieving the highest fidelity at the end of the simulation.   

Finally, we investigate the effect of two alternative definitions of the cost function on the shot-allocation strategy. Typically, the cost function is defined as the collective variance of one or more quantities, which goes to zero in the limit of an infinite number of shots. When the total number of shots is fixed, one attempts to distribute the shots in a way that minimizes the cost function. In Eq.~\eqref{eq:variance}, the collection of quantities, whose total variance one aims to minimize, is the variational parameter vector $\bth$, the solution to the equations of motion. The overall goal for noise mitigation in a dynamical simulation is to make the next-step wavefunction as close to the statevector as possible. Thus, it seems reasonable to define the cost function as the total variance of the next-step wavefunction $\ket{\Psi(t+\delta t)}$. Additionally, the equation of motion Eq.~\eqref{eq:eom} is established by minimizing the McLachlan distance $L^2$ during each time step. The minimal value of $L^2$ is essential in monitoring the performance of the variational ansatz. This is especially important in adaptive algorithms~\cite{VQITE, gomesAdaptiveVariationalQuantum2021}, since $L^2$ is explicitly needed to determine whether an adaptive process for expanding the ansatz is necessary. Thus, in some applications, one can also define the cost function as the variance of $L^2$, so that $L^2$ can be measured as accurately as possible given a shot budget. 
All cost function definitions share similar structures as Eq.~\eqref{eq:variance}, with the only difference being the definitions of $p_{\kappa}$. In Appendix~\ref{appendix:a}, we present the definition of these cost functions and the analytical expressions of $p_{\kappa}$. Algorithm~\ref{alg:shotAlloc}
is independent of a specific definition of the cost function. 

\begin{figure}[t]
    \centering
    \includegraphics[width=\linewidth]{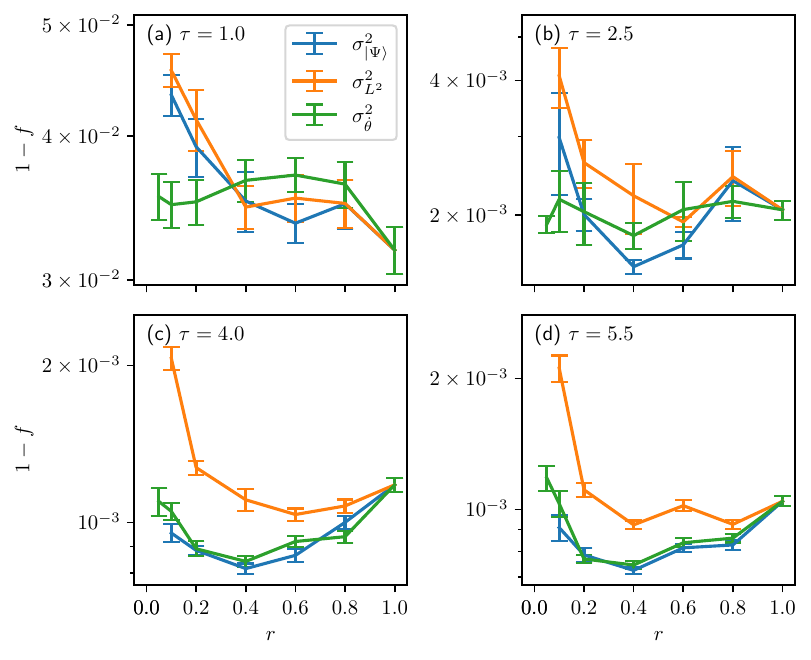}
    \caption{
    The infidelity as a function of $r$ for several instants during the imaginary time evolution for noisy VQITE simulations of the critical TFIM with $N=6$. The shot allocation algorithm is implemented with a few different definitions of the FOM, including $\sigma_{\dot{\bth}}^2$,
    $\sigma_{L^2}^2$, and $\sigma_{\ket{\Psi}}^2$.
    }
    \label{fig:N6-f-r-all}
\end{figure}

Figure~\ref{fig:N6-f-r-all} shows the average infidelity as a function of uniform budget fraction $r$ at different times during the imaginary time evolution. The three cost functions as described above were used in the shot allocation algorithm. 50 independent runs were carried out for each cost function. Similar to Fig.~\ref{fig:N6-f-r}, the infidelity at the early stage ($\tau=1$) does not indicate an advantage of the shot allocation strategy. At later stages for times $\tau=4$ and 5.5, however, shots allocated according to the cost functions $\sigma_{\dot{\bth}}^2$ and $\sigma_{\ket{\Psi}}^2$ both show optimal infidelity at $r=0.4$, much improved from the uniform shot allocation $r=1.0$. On the other hand, the cost function $\sigma_{L^2}^2$ is ineffective at all time steps, possibly because the dynamical evolution in VQITE does not explicitly depend on the optimized value of $L^2$. Future work should explore the use of $\sigma_{L^2}$ in adaptive VQDS such as AVQITE~\cite{AVQITE}. Since the next-step wavefunction $\ket{\Psi(t+\delta t)}$ can be expressed in terms of $\dot{\bth}$ (see Appendix~\ref{appendix:a}), $\sigma_{\ket{\Psi}}^2$ is essentially a reweighting of the variances in all the elements of $\dot{\bth}$, as compared with Eq.~\eqref{eq:variance}, in which the elements of $\dot{\bth}$ are weighted equally. The close performance of these two cost functions in the late stages [see Fig.~\ref{fig:N6-f-r-all} (c) and (d)] indicates that such reweighting process, while introducing extra computational cost, is largely unnecessary.

\section{Conclusions}
\label{sec:conclusions}
In this work, we studied the effects of finite sampling noise on variational quantum dynamics simulations based on McLachlan’s variational principle, with particular emphasis on imaginary-time evolution for ground-state preparation. Using transverse-field Ising chains as benchmark models, we demonstrated that sampling noise can severely degrade the performance of VQITE due to the ill-conditioned nature of the quantum geometric tensor entering the equations of motion. Among the stabilization techniques considered, we found that Tikhonov regularization offers a favorable balance between numerical stability and robustness against shot noise in noisy simulations.

Beyond matrix regularization, we addressed the critical question of how to optimally distribute a finite measurement budget across the quantum circuits required to evaluate the variational equations of motion. We introduced a general shot-allocation strategy based on minimizing a cost function characterizing noise-induced errors. Going beyond previous works~\cite{PRXQuantum.2.030324}, we show that maintaining a sufficient (minimal) number of shots for each measurement circuit is critical to obtain reliable convergence in VQITE simulations. Our results show that allocating measurements according to a cost function defined as the collective variance of the solution to the equations of motion consistently outperforms uniform shot distributions. With a necessary minimal number of shots of about 40\% of the average number of shots for each measurement, this approach yields substantial improvements in state fidelity and enables a significant reduction in the total number of measurements required to reach a given accuracy.

To conclude, our findings highlight that careful noise-aware design of both numerical solvers and measurement strategies is essential for realizing the full potential of variational quantum dynamics on NISQ devices. The shot-allocation framework presented here is general and can be readily applied to other variational algorithms, including real-time VQDS, adaptive variational schemes, and quantum natural-gradient-based optimization methods. We expect that these techniques will play an important role in enabling scalable and measurement-efficient quantum simulations on near-term quantum hardware.

\acknowledgments
This work was supported by the U.S. Department of Energy (DOE), Office of Science, Basic Energy Sciences, Materials Science and Engineering Division, including the grant of computer time at the National Energy Research Scientific Computing Center (NERSC) in Berkeley, California. The research was performed at the Ames National Laboratory, which is operated for the U.S. DOE by Iowa State University under Contract No. DE-AC02-07CH11358.
PPO acknowledges support by the German Federal Ministry of Research, Technology and Space via the Research Program Quantum Systems with contract number 13N17150 during the final stages of this work.

%

\newpage
\onecolumngrid
\appendix
\section{Cost function and the derivative with respect to raw measurements}
\label{appendix:a}
A shot-allocating strategy is determined by minimizing a cost function ($C$) subject to the constraint that the total number of shots is fixed. Generally, the cost function is the variance of a quantity or the sum of the variances of various quantities that can be derived from the raw measurements: 
\be\label{eq:cost}
C=\sum_{j=1}^{N_Q}\sigma_{Q_j}^2 \,\, (N_Q\ge 1). 
\ee
$\sigma_{Q_j}^2$ is determined using the chain rule for variance propagation:
\be\label{eq:chain}
\sigma_{Q_j}^2=\sum_\kappa\left|\pdv{Q_j}{m_\kappa}\right|^2\frac{\sigma_{m_\kappa}^2}{\mathcal{M}_\kappa},
\ee
where $m_\kappa$ are independent raw measurements, whose intrinsic variance due to the quantum nature is calculated as $\sigma_{m_\kappa}^2=\langle m_\kappa^2\rangle-\langle m_\kappa\rangle^2$, and $\mathcal{M}_\kappa$ is the number of measurements used to measure $m_\kappa$. For the imaginary time evolution of the TFIM, measurements of $\mathcal{D}_{\mu \nu}$ and $\mathcal{O}_\mu$ are made for calculating $M_{\mu \nu}$ according to $M_{\mu\nu}=\Re(\mathcal{D}_{\mu\nu}+\mathcal{O}_\mu\mathcal{O}_\nu)$, and measurements of energy in $Z$- and $X$-basis ($E=E_Z+E_X$) and $V$ include are made for calculating $V_\mu$ using the parameter shift rule $V_\mu=(E_\mu^+-E_\mu^-)/4$ where $E_\mu^+$ and $E_\mu^-$ stand for $E(\theta_\mu+\pi/2)$ and $E(\theta_\mu-\pi/2)$, respectively. We first calculate the derivative of $C$ with respect to $M_{\mu \nu}$ and $V_\mu$; then, the derivative with respect to the raw measurements can be easily reached:
\begin{align}
    \pdv{Q_j}{\mathcal{D}_{\mu \nu}} & = \pdv{Q_j}{M_{\mu \nu}},\label{eq:pdv_D}\\
    \pdv{Q_j}{\mathcal{O}_\mu } & = \sum_\gamma{\left(\pdv{Q_j}{M_{\mu \gamma}}+\pdv{Q_j}{M_{\gamma \mu}}\right)\mathcal{O}_\gamma},\label{eq:pdv_O}\\
    \pdv{Q_j}{E_\mu^+} & =-\pdv{Q_j}{E_\mu^-}=-\frac{1}{4}\pdv{Q_j}{V_\mu}.\label{eq:pdv_E}
\end{align}
Below, we provide a few examples of the cost function and its derivative.
\subsection{Variance of $\dot{\bth}$}
Eq.~\ref{eq:variance} defines the cost function as the sum of the variances of all the elements of $\dot{\bth}$, the solution to the equations of motion in Eq.~\ref{eq:eom}: $C=\sum_\mu\sigma_{\dot{\theta}_\mu}^2$. Since $\dot{\theta}_\mu=\sum_\nu M_{\mu \nu}^{-1}V_\nu$, we have:
\begin{equation}\label{eq:pdv_theta_v}
\pder{\dot{\theta}_\mu}{V_\alpha} = \sum_{\nu} M^{-1}_{\mu \nu} \pder{V_\nu}{V_\alpha} = M^{-1}_{\mu \alpha}\, ,
\end{equation}
and
\begin{equation}\label{eq:pdv_theta_M}
\pder{\dot{\theta}_\mu}{M_{\alpha \beta}} = \sum_{\nu} \pder{M^{-1}_{\mu \nu}}{M_{\alpha \beta}} V_\nu = -\sum_{\nu \mu' \nu'} M^{-1}_{\mu \nu'} \pder{M_{\nu' \mu'}}{M_{\alpha \beta}} M^{-1}_{\mu' \nu} V_\nu = -\sum_{\nu} M^{-1}_{\mu \alpha} M^{-1}_{\beta \nu} V_{\nu}=M^{-1}_{\mu \alpha}\dot{\theta}_\beta\,.
\end{equation}
The second equality here occurs because $\pder{(M^{-1} M)}{M_{\mu \nu}} = \pder{M^{-1}}{M_{\mu\nu}}M + M^{-1}\pder{M}{M_{\mu\nu}} = 0$. Eq.~\ref{eq:pdv_theta_M} is in general asymmetric with respect to $\alpha$ and $\beta$. In practice, the matrix $M$ is kept symmetric during the dynamical simulation, that is, the updates on $M_{\alpha\beta}$ and $M_{\beta\alpha}$ are the same; thus, we can have the following symmetrized form:
\be\label{eq:pdv_theta_M_symm}
\pder{\dot{\theta}_\mu}{M_{\alpha \beta}}=M^{-1}_{\mu \alpha}\dot{\theta}_\beta + M^{-1}_{\mu \beta}\dot{\theta}_\alpha.
\ee

\subsection{Variance of $\ket{\Psi(t+\delta t)}$}
We can write the update rule of the variational parameters as $f_{\mu}[t] = \theta_{\mu}[t] + \delta \theta_{\mu}[t]$ for $\delta \theta_{\mu}[t] = \sum_{\nu} M^{-1}_{\mu \nu} V_{\nu} \delta t$ and use this to write one time step of the variational ansatz, Eq.~\eqref{eq:ansatz}, as
\begin{equation}
\ket{\Psi(\boldsymbol{\theta}[t + \delta t])} = \prod_{\mu = 1}^{N_\theta} e^{-i P_{\mu} f_{\mu}[t]} \ket{\psi_0}\, ,
\end{equation}
where note that we have dispensed with the layered form of the ansatz (see Eq.~\ref{eq:ansatz}) and now $\mu = 1, \dots, N_\theta$, i.e., $\mu$ runs over every variational parameter. The $P_\mu$ remain the Pauli strings associated with variational parameters $\theta_\mu$, as in the main text. Then the derivative with respect to a variational parameter is:
\begin{equation}
\pder{\ket{\Psi(\boldsymbol{\theta}[t + \delta t])}}{\theta_{\nu}} = \prod_{\mu > \nu} e^{-i P_\mu f_\mu[t]} \round{-i P_{\nu}} \prod_{\mu \leq \nu} e^{-i P_\mu f_\mu[t]}\ket{\psi_0}\, .
\end{equation}
We can define the cost function as:
\begin{equation}
    \sigma_{\ket{\Psi}}^2 = \sum_\kappa \frac{\partial\bra{\Psi(t+\delta t)}}{\partial m_\kappa}\frac{\partial\ket{\Psi(t+\delta t)}}{\partial m_\kappa}\frac{\sigma_{m_\kappa}^2}{\mathcal{M}_\kappa}.
\end{equation}\label{eq:fid_cost} 
The derivative of  $\ket{\Psi(\boldsymbol{\theta}[t + \delta t]}$ with respect to $V_\alpha$ is:
\begin{align}
\pder{\ket{\Psi(\boldsymbol{\theta}[t + \delta t])}}{V_\alpha} &= \sum_{\nu} \prod_{\mu > \nu} e^{-i P_\mu f_\mu[t]} \round{-i P_{\nu} \pder{f_{\nu}[t]}{V_\alpha}} \prod_{\mu \leq \nu} e^{-i P_\mu f_\mu[t]} \ket{\psi_0} \nn \\
		&= \sum_{\nu} \prod_{\mu > \nu} e^{-i P_\mu f_\mu[t]} \round{-i P_{\nu} M^{-1}_{\nu \alpha} \delta t} \prod_{\mu \leq \nu} e^{-i P_\mu f_\mu[t]}\ket{\psi_0} \nn \\
		&= \sum_{\nu} M^{-1}_{\nu \alpha} \delta t ~\pder{\ket{\Psi(\boldsymbol{\theta}[t + \delta t])}}{\theta_{\nu}}\, .
\end{align}
The derivative with respect to $M_{\alpha \beta}$ is:
\begin{align}
\pder{\ket{\Psi(\boldsymbol{\theta}[t + \delta t])}}{M_{\alpha \beta}} &= \sum_{\nu} \prod_{\mu > \nu} e^{-i P_\mu f_\mu[t]} \round{-i P_{\nu} \pder{f_{\nu}[t]}{M_{\alpha \beta}}} \prod_{\mu \leq \nu} e^{-i P_\mu f_\mu[t]} \ket{\psi_0} \nn \\
		&= -\sum_{\nu} \prod_{\mu > \nu} e^{-i P_\mu f_\mu[t]} \round{-i P_{\nu}} \square{\round{M^{-1}_{\nu \alpha} \dot{\theta}_\beta+M^{-1}_{\nu \beta} \dot{\theta}_\alpha} \delta t} \prod_{\mu \leq \nu} e^{-i P_\mu f_\mu[t]}\ket{\psi_0} \nn \\
		&= -\sum_{\nu} \round{M^{-1}_{\nu \alpha} \dot{\theta}_\beta+M^{-1}_{\nu \beta} \dot{\theta}_\alpha} \delta t ~\pder{\ket{\Psi(\boldsymbol{\theta}[t + \delta t])}}{\theta_{\nu}}\, .
\end{align}
In these forms, the inner products are easily computed:
\begin{align}
\pder{\bra{\Psi(\boldsymbol{\theta}[t + \delta t])}}{V_\alpha} \pder{\ket{\Psi(\boldsymbol{\theta}[t + \delta t])}}{V_\alpha} & = \sum_{\nu \nu'} M^{-1}_{\alpha \nu} M^{-1}_{\nu' \alpha} \delta t^2 \pder{\bra{\Psi(\boldsymbol{\theta}[t + \delta t])}}{\theta_{\nu}} \pder{\ket{\Psi(\boldsymbol{\theta}[t + \delta t])}}{\theta_{\nu'}} \nn \\
			& = \sum_{\nu \nu'} M^{-1}_{\alpha \nu} M^{-1}_{\nu' \alpha} \delta t^2 \SD_{\nu \nu'} \nn \\
			& = \sum_{\nu \nu'} M^{-1}_{\alpha \nu} M^{-1}_{\nu' \alpha} \delta t^2 ~\rm{Re}\curly{\SD_{\nu \nu'}}\, , \label{eq:Vmu}
\end{align}
where the first equality is simply performing the Hermitian conjugate and multiplying; the second equality occurs because of the fact that $M_{\mu \nu}$ is real and symmetric, and by the definition of $\SD_{\mu \nu}$. The final equality occurs due to the summations and the Hermiticity of $\SD$ as already mentioned | i.e., for every matrix element there is always a corresponding matrix element that cancels the imaginary component. The equation for the other inner product proceeds similarly:
\begin{align}
\pder{\bra{\Psi(\boldsymbol{\theta}[t + \delta t])}}{M_{\alpha \beta}} \pder{\ket{\Psi(\boldsymbol{\theta}[t + \delta t])}}{M_{\alpha \beta}} & = \sum_{\nu \nu'} \round{\dot{\theta}_\beta M^{-1}_{\alpha \nu} +  \dot{\theta}_\alpha M^{-1}_{\beta \nu}} \round{M^{-1}_{\nu' \alpha} \dot{\theta}_\beta+M^{-1}_{\nu' \beta} \dot{\theta}_\alpha} \delta t^2 \pder{\bra{\Psi(\boldsymbol{\theta}[t + \delta t])}}{\theta_{\nu}} \pder{\ket{\Psi(\boldsymbol{\theta}[t + \delta t])}}{\theta_{\nu'}} \nn \\
			& = \sum_{\nu \nu'} \round{\dot{\theta}_\beta M^{-1}_{\alpha \nu} +  \dot{\theta}_\alpha M^{-1}_{\beta \nu}} \round{M^{-1}_{\nu' \alpha} \dot{\theta}_\beta+M^{-1}_{\nu' \beta} \dot{\theta}_\alpha} \delta t^2 \SD_{\nu \nu'} \nn \\
			& = \sum_{\nu \nu'} \round{\dot{\theta}_\beta M^{-1}_{\alpha \nu} +  \dot{\theta}_\alpha M^{-1}_{\beta \nu}} \round{M^{-1}_{\nu' \alpha} \dot{\theta}_\beta+M^{-1}_{\nu' \beta} \dot{\theta}_\alpha} \delta t^2~\rm{Re}\curly{\SD_{\nu \nu'}}\, . \label{eq:Mmunu}
\end{align}
Considering the symmetry in the inverse matrix $M^{-1}$, that is, $M^{-1}_{\mu \nu} = M^{-1}_{\nu \mu}$, for a given pair of $(\alpha, \beta)$, the imaginary component cancels as well between the $(\nu,\nu')$ and $(\nu',\nu)$ terms. $\delta t^2$ in both Eq.~\ref{eq:Vmu} and~\ref{eq:Mmunu} can be treated as an overall scaling factor, which does not affect the shot allocation. 

\subsection{Variance of $L^2$}
Using $\dot{\theta}_\mu=\sum_\nu M_{\mu \nu}^{-1}V_\nu$, the minimized $L^2$ can be written as (see Eq.~\ref{L2}):
\be
L^2 = -\sum_{\mu \nu} V_\mu M^{-1}_{\mu \nu} V_\nu + 2 \text{var}_{\theta}H\, .
\ee
The last term stands for the variance of the energy, which does not depend on $M$ or $V$.
The derivative with respect to $V_\alpha$:
\be\label{eq:pder_L2_V}
\pder{L^2}{V_\alpha}=-2\sum_\mu M^{-1}_{\alpha \mu}V_\mu = -2\dot{\theta}_\alpha\, .
\ee
The derivative with respect to $M_{\alpha \beta}$:
\be\label{eq:pder_L2_M}
\pder{L^2}{M_{\alpha \beta}} = -\sum_{\mu \nu} V_\mu \pder{M^{-1}_{\mu \nu}}{M_{\alpha \beta}} V_\nu = -\sum_{\mu \nu} V_\mu \round{M^{-1}_{\mu \alpha}M^{-1}_{\beta \nu}+M^{-1}_{\mu \beta}M^{-1}_{\alpha \nu}} V_\nu=-2\dot{\theta}_\alpha \dot{\theta}_\beta\, .
\ee
Similar to Eq.~\ref{eq:pdv_theta_M_symm}, Eqs.~\ref{eq:Mmunu} and ~\ref{eq:pder_L2_M} also reflect the fact that $M_{\alpha \beta}$ and $M_{\beta \alpha}$ are varied simultaneously during the simulation.
\end{document}